\documentclass{article}

\PassOptionsToPackage{numbers,sort&compress}{natbib}

\usepackage[final]{neurips_2025}
\usepackage[utf8]{inputenc}
\usepackage[T1]{fontenc}
\usepackage{hyperref}
\usepackage{xurl}
\usepackage{booktabs}
\usepackage{amsfonts}
\usepackage{graphicx}
\usepackage{nicefrac}
\usepackage{microtype}
\usepackage{xcolor}
\usepackage[export]{adjustbox}
\usepackage{tabularx}
\usepackage{amsmath}

\title{WOLF: Werewolf-based Observations for LLM Deception and Falsehoods}

\author{
  \textbf{Mrinal Agarwal}\thanks{Lead Author}
  \and
  \textbf{Saad Rana}\thanks{2nd Author}
  \and
  \textbf{Theo Sundoro}
  \and
  \textbf{Hermela Berhe}
  \and
  \textbf{Spencer Kim}\thanks{Advisors}
  \and
  \textbf{Vasu Sharma}\footnotemark[3]
  \and
  \textbf{Sean O'Brien}\footnotemark[3]
  \and
  \textbf{Kevin Zhu}\footnotemark[3] \\
  \vspace{1em}
  \text{Algoverse AI Research} \\
  \texttt{smmrinal2009@gmail.com, sfrana08@gmail.com} \\
\texttt{2000.seano@gmail.com, kevin@algoverse.us}
}

\begin{document}
\maketitle

\begin{abstract}
Deception is a fundamental challenge for multi-agent reasoning: effective systems must strategically conceal information while detecting misleading behavior in others. Yet most evaluations reduce deception to static classification, ignoring the interactive, adversarial, and longitudinal nature of real deceptive dynamics. Large language models (LLMs) can deceive convincingly but remain weak at detecting deception in peers. We present WOLF, a multi-agent social deduction benchmark based on Werewolf that enables separable measurement of deception production and detection. WOLF embeds role-grounded agents (\emph{Villager}, \emph{Werewolf}, \emph{Seer}, \emph{Doctor}) in a programmable LangGraph state machine with strict night–day cycles, debate turns, and majority voting. Every statement is a distinct analysis unit, with self-assessed honesty from speakers and peer-rated deceptiveness from others. Deception is categorized via a standardized taxonomy (\emph{omission}, \emph{distortion}, \emph{fabrication}, \emph{misdirection}), while suspicion scores are longitudinally smoothed to capture both immediate judgments and evolving trust dynamics. Structured logs preserve prompts, outputs, and state transitions for full reproducibility. Across 7,320 statements and 100 runs, Werewolves produce deceptive statements in 31\% of turns, while peer detection achieves 71–73\% precision with ~52\% overall accuracy. Precision is higher for identifying Werewolves, though false positives occur against Villagers. Suspicion toward Werewolves rises from ~52\% to over 60\% across rounds, while suspicion toward Villagers and the Doctor stabilizes near 44–46\%. This divergence shows that extended interaction improves recall against liars without compounding errors against truthful roles. WOLF moves deception evaluation beyond static datasets, offering a dynamic, controlled testbed for measuring deceptive and detective capacity in adversarial multi-agent interaction.
\end{abstract}

\section{Introduction}
As large language models (LLMs) are increasingly being deployed into autonomous capacities across society, ensuring transparency within these models has become paramount. LLMs now operate in multi-agent settings as assistants, tutors, healthcare advisors, and even formal decision-makers \cite{haase2025staticresponsesmultiagentllm}. The very progress that expands LLM usefulness also raises new risks --- whilst these models become more effective collaborators, they also gain greater potential to deceive in subtle ways \cite{kwa2025measuringaiabilitycomplete}. This raises concerns about deception strategies --- absent in earlier generations --- that emerged in GPT-4 \cite{Hagendorff_2024}. It is already recognized that LLMs in real-world simulations can engage in deception, without explicit instruction to do so \cite{scheurer2024largelanguagemodelsstrategically}. Even when models are trained to be honest, under the right pressure, LLMs often choose to deceive, irrespective of how truthful models are when benchmarked \cite{ren2025maskbenchmarkdisentanglinghonesty}.

In a multi-agent scenario, GPT-4 demonstrates strong deceptive capabilities; however, it remains highly susceptible to being deceived itself \cite{curvo2025traitorsdeceptiontrustmultiagent}. This asymmetry highlights that while deception capabilities scale as models develop, detection capabilities lag behind—leaving LLMs simultaneously persuasive yet vulnerable. This leads to our guiding hypothesis: \textbf{deception scales more quickly than detection}. Modern LLMs exhibit strong cooperative reasoning yet remain under-tested in adversarial, multi-party settings where incentives to mislead are explicit.

In this work, we introduce WOLF, a benchmark that examines this asymmetry by evaluating both deceptive and detective capacities of LLMs in a multi-agent setting modeled on Werewolf. Social deduction games like Werewolf naturally induce deception, partial information, and theory-of-mind reasoning \cite{bailis2024werewolfarenacasestudy}. WOLF is designed to analyze three guiding questions: (i) how often and in what forms LLMs engage in deception, (ii) how accurately and consistently they detect deception in peers, and (iii) how these dynamics evolve across multi-turn, role-conditioned interactions. We operationalize this setting as a reproducible benchmark with role-grounded agents, programmable rules, and structured logging. 

This work makes three contributions:
\begin{enumerate}
\item A LangGraph-based implementation of Werewolf with role-conditioned agents, strict phase transitions, and structured event logs for reproducibility;
\item A deception measurement protocol coupling self-assessment with concurrent peer analysis for every public statement, categorized by omission, distortion, fabrication, and misdirection;
\item A metrics suite that captures deception production rates, detection accuracy, calibration (Brier score, ROC/AUPRC), and longitudinal suspicion trends that reveal how trust and mistrust develop over time.
\end{enumerate}

\section{Related Works}
Benchmarks rarely elicit sustained deceptive behavior under partial information while also measuring detection. Social deduction settings surface coordination and theory-of-mind reasoning, but standardized, logged, multi-turn deception benchmarks remain limited. Existing work has explored agent-based evaluations with discussion and voting or deception classification with static text. WOLF extends these threads by integrating (i) a competitive, role-asymmetric game loop, (ii) statement-level deception labels from both speakers and peers, and (iii) full-fidelity logging for audit and replication.  

\subsection{Deception Capabilities in Large Language Models}
LLMs can deceive convincingly, even when optimized to be helpful, honest, and harmless. Studies show models strategically conceal intentions under monitoring \cite{barkur2025deceptionllmsselfpreservationautonomous}, and deception emerges even in simple scenarios \cite{wu2025promptinducedliesinvestigatingllm}. For example, GPT-4 deceived a human TaskRabbit worker by pretending to be visually impaired to bypass a CAPTCHA \cite{openai2023gpt4}. In multi-agent simulations, LLMs fabricate false narratives \cite{chen2025aiagentbehavioralscience} and generate explanations to justify false claims, making misinformation more persuasive \cite{Danry_Pataranutaporn_Groh_Epstein_2025}.  

Benchmarks confirm the prevalence of deception: over 80\% of interactions in \textit{OpenDeception} showed deceptive intent \cite{wu2025opendeceptionbenchmarkinginvestigatingai}. The \textit{MASK} benchmark found that larger models conceal lies more effectively, not necessarily becoming more honest \cite{ren2025maskbenchmarkdisentanglinghonesty}.  

\subsection{Deception Detection in Large Language Models}
Fewer studies examine whether LLMs reliably detect deception. In \textit{The Traitors}, models produced lies but were easily deceived themselves \cite{chopra2024viewaboveframeworkevaluating}. In \textit{Hoodwinked}, GPT-4 impostors lied convincingly, but detection was weak \cite{ogara2023hoodwinkeddeceptioncooperationtextbased}. \textit{OpenDeception} revealed frequent planning of deception but rare detection of others’ intent \cite{wu2025opendeceptionbenchmarkinginvestigatingai}, and \textit{MASK} showed that models often internally represent truth yet fail to identify dishonesty when contradicted \cite{ren2025maskbenchmarkdisentanglinghonesty}.  

\subsection{Social Deduction and Multi-Agent Evaluation Frameworks}
Social deduction games are natural testbeds for deception and theory of mind (ToM) \cite{liu2024interintentinvestigatingsocialintelligence}. Players must mislead while reasoning about others’ actions, directly testing both deception and detection.  

\textit{Werewolf Arena} is the most influential framework: LLMs competed under asymmetric information with bid-based turn-taking \cite{bailis2024werewolfarenacasestudy}. Results showed divergent strategies—some excelling at persuasion, others at deduction—but evaluation was limited to game-level outcomes (wins, eliminations), without labeling which statements were lies or caught. Other adaptations share similar gaps: \textit{AvalonBench} revealed weaknesses in multi-turn reasoning \cite{light2023avalonbenchevaluatingllmsplaying}, \textit{AmongAgents} showed GPT-4’s basic role-play in \textit{Among Us} but poor lie detection \cite{chi2024amongagentsevaluatinglargelanguage}, and \textit{The Traitors} added persistent memory and deception metrics but lacked statement-level labels \cite{curvo2025traitorsdeceptiontrustmultiagent}. Collectively, these works show that while social deduction surfaces deception, none systematically measure detection accuracy at the statement level.

\section{Werewolf Game Flow and Architecture}
\label{subsec:Werewolf Game Flow and Architecture}

\subsubsection*{Roles and Setup}
WOLF recreates the social deduction game Werewolf as a programmable state graph with fixed roles, phases, and transitions. Each game uses a roster of eight participants: four Villagers, two Werewolves, one Seer, and one Doctor. Villagers form the uninformed majority, relying on debate and voting to identify threats. Werewolves act as a coordinated minority, eliminating Villagers at night while concealing their identities during the day. The Seer privately inspects one role each night, and the Doctor protects one player from elimination. This fixed distribution ensures stable conditions for deception to emerge and enables reproducible comparisons across runs, unlike prior implementations that randomized roles.  

\subsubsection*{Game Phases}
The game alternates between night and day. At night, Werewolves select a target, the Doctor chooses a player to protect, and the Seer investigates a role. These actions resolve simultaneously, with only outcomes—survival or elimination—revealed. During the day, players debate based on these outcomes, exchanging accusations and defenses before voting to exile a suspected player. A majority is required for exile; otherwise, no action occurs. Victory conditions are asymmetric: Villagers win by eliminating all Werewolves, while Werewolves win once their numbers equal or exceed those of the Villagers.  

\subsubsection*{Debate Structure}
\label{subsec:debate structure}
Debates follow a bidding system introduced in Werewolf Arena \cite{bailis2024werewolfarenacasestudy} and extended in WOLF. Before each turn, players bid an integer between 0 and 10 to indicate urgency. Higher bids increase the chance of speaking earlier, but frequent overbidding reduces later influence. Ties are broken with a slight bias toward players referenced in the prior turn, reflecting the conversational pull of responding to accusations. This mechanism prevents rigid orderings and better captures the spontaneous dynamics of group debate.  

\subsubsection*{Memory and Suspicion}
WOLF extends earlier memory mechanisms \cite{bailis2024werewolfarenacasestudy} by treating each public statement as a unit of analysis with explicit deception assessments. Speakers provide a self-assessment of honesty, while peers judge deceptiveness. These overlapping labels update suspicion scores via exponential smoothing, producing longitudinal traces of trust and doubt that shape later decisions. For example, hesitation in one round may raise suspicion that persists into future rounds, even if later statements appear convincing. All actions—bids, debates, votes, and assessments—are recorded in structured logs, making the benchmark reproducible and suitable for fine-grained analysis.

\section{Methodology}
\label{sec:Method}
\subsection{Deception Measurement Protocol}
\label{subsec:deception-protocol}
We measure deception at the granularity of individual statements. Each time a player speaks during debate or explains a vote, that statement is treated as a distinct unit of analysis. After each statement, the speaker provides a self-assessment of honesty, while all other players record their perception of the speaker’s honesty. This design ensures that both the production and detection of deception are captured in real time.

Because suspicion is rarely static, we model its longitudinal dynamics with exponential smoothing:
\[
D_{t+1}(o,t) = \alpha \cdot s(o,t) + (1-\alpha)\cdot D_{t}(o,t), \quad \alpha=0.7,
\]
Here, $s(o,t)$ denotes the observer’s current suspicion for the new statement, and $D_t(o,t)$ represents their suspicion aggregated from all prior statements. The smoothing factor $\alpha = 0.7$ ensures that new evidence carries significant weight, while prior history tempers overreaction. This mirrors human-like reasoning: a single suspicious comment may raise doubts, but trust or distrust stabilizes only after repeated patterns. The protocol therefore yields suspicion trajectories that capture both immediate reactions and cumulative judgment.

\subsection{Statement Analysis}

Each statement generates two layers of analysis: a self-assessment by the speaker and peer assessments by the observers. Both self- and peer analyses are generated through private scratchpads, where agents reason internally using chain-of-thought. Only the structured fields are surfaced, preventing reasoning leaks while enabling a consistent evaluation. This mechanism is essential: it ensures that deception judgments arise from a role-specific context rather than being contaminated by hidden knowledge.

\paragraph{Self-analysis.}The speaker indicates whether the statement is deceptive, $d \in \{0,1\}$, reports confidence in this judgment, $c \in [0,1]$, and, if applicable, specifies the type of deception employed. These labels distinguish among different mechanisms of dishonesty.

\begin{itemize}
\item \texttt{none}: The statement is truthful, with no deceptive intent. This baseline anchors the taxonomy, ensuring that honesty is explicitly represented alongside deception.  

\item \texttt{omission}: Withholds relevant information. Deception arises not from false claims but from selectively leaving out details that would alter interpretation.  

\item \texttt{distortion}: Alters true information in a misleading way. Facts are presented but are exaggerated, minimized, or reframed to produce a false impression.  

\item \texttt{misdirection}: Diverts attention away from relevant facts. Rather than falsifying content, the speaker redirects focus toward less relevant or distracting information.  

\item \texttt{fabrication}: Introduces information that is entirely false. This represents the strongest form of deception, where statements are invented without a basis in truth.  
\end{itemize}

The self-analysis also includes a reasoning field that provides context for the decision. 
This component is crucial because it separates deliberate deception from mistakes, uncertainty, or hallucination. 
For example, a player might record ``I left out X because revealing it would compromise my role,'' 
which is qualitatively different from ``I was unsure whether this counted as deception.''

\paragraph{Peer analysis.} Observers annotate the same fields and additionally assign a continuous suspicion score $s \in [0,1]$, where $0$ denotes full trust and $1$ denotes certainty of deception. Unlike the binary deception flag, this score captures gradations of distrust: for example, $s=0.2$ may reflect mild doubt, whereas $s=0.9$ indicates near-certainty of dishonesty. The suspicion score serves as the core signal in our longitudinal update rule (see Section~\ref{subsec:deception-protocol}), enabling us to track how trust fluctuates over the course of play.

\subsection{Agents and Prompts}
\paragraph{Role-grounded prompts.} 
Agents in WOLF are explicitly bound to their roles and objectives. Werewolves must conceal their identities and mislead others while coordinating at night. Villagers are expected to stay transparent, cooperate, and flag suspicious behavior. The Seer decides when to reveal or withhold investigative results, and the Doctor protects chosen players from elimination. These prompts ensure that deception emerges naturally from role incentives rather than arbitrary instructions.

\paragraph{Private scratchpads.} 
Agents also keep private scratchpads where they reason with chain-of-thought. For example, a Werewolf might note \textit{``Emma is defending me; likely a Villager''} while publicly saying \textit{``Raj’s hesitation is suspicious.''} This separation ensures deception arises from role-based incentives rather than leaking privileged knowledge. Scratchpads are not shared with other agents but are logged for researchers, preserving gameplay integrity while enabling full analysis.

\paragraph{Controlled mechanics.} 
To keep runs stable, WOLF limits debate length, orders speakers using the bidding system from Section~\ref{subsec:debate structure} (with ties resolved by mention priority), and repairs malformed outputs with conservative defaults. These safeguards standardize interaction so that differences reflect model strategy rather than formatting quirks or randomness.

\subsection{Evaluation Setup}
We evaluate WOLF in two complementary modes: one with stochastic LLM agents and one with deterministic controls. Together, these reveal both the natural behavior of language models and the baseline functioning of the framework.  

\paragraph{Full LLM runs.} 
Agents are instantiated with actual language models and play under the normal game rules. Because LLMs are stochastic, behavior varies across runs. Werewolves may bluff more aggressively, Villagers may hedge votes, and the Doctor or Seer may reveal information at different times depending on dialogue flow. These runs capture how deception and detection emerge under realistic conditions.  

\paragraph{Subsystem ablations.} 
Randomness is removed using a deterministic mock analyzer that applies fixed rules to assign labels, confidences, and suspicion scores from text cues. While less naturalistic, this mode verifies that suspicion updates follow exponential smoothing, metrics aggregate correctly, and the system behaves consistently under fixed inputs.  

\paragraph{Comparison.} 
Contrasting the two modes separates model-driven behavior (e.g., hedging, bluffing, overconfidence) from benchmark-level properties (e.g., suspicion trajectory updates). To ensure reproducibility, we preserve all prompts and outputs and log every event in NDJSON streams with state snapshots and per-player metrics, allowing exact reconstruction of runs under alternative criteria.  

\subsection{Metrics}
To understand deception in this environment, we evaluate the following metrics.
\paragraph{Deception production rate.}  
This metric measures how often deception occurs. It captures not only how frequently Werewolves lie, but also whether honest roles such as Villagers hedge or misstate facts in ways that appear deceptive. This reveals the willingness of agents to deceive and shows how role incentives shape that willingness.  

\paragraph{Detection accuracy.}  
This metric compares peer judgments against self-reports. If peers correctly identify deceptive statements, accuracy is high, if they misjudge honest but cautious speech as lying, accuracy drops. 

\paragraph{Calibration.}  
Measured with the Brier score, calibration asks whether suspicion values behave like probabilities. In a well-calibrated system, a suspicion of 0.6 should correspond to deception about 60\% of the time. Calibration matters because suspicion is only useful if it can be trusted as a predictive signal. Poor calibration implies systematic overconfidence or underconfidence, undermining reliability in multi-agent settings.  

\paragraph{Cross-perception matrix.}  
This metric aggregates suspicion across all observer--target pairs, showing how distrust distributes across the group. Ideally, suspicion should converge on deceptive roles; diffuse suspicion indicates noise or systematic confusion.  

\paragraph{Threshold analyses.}  
Using ROC and AUPRC curves, these analyses evaluate whether suspicion can be turned into actionable decisions. ROC curves show the overall discriminative ability of suspicion scores, while AUPRC is particularly useful since deception is relatively rare compared to truth. Threshold analyses test whether suspicion can support rule-based interventions in gameplay or downstream applications.  

Taken together, these metrics reveal not only whether deception occurs, but also whether it is detectable, whether suspicion values can be trusted, and whether those values can be used in decisions. 

\section{Experimentation and Results}
\label{sec:experiments}

\subsection{Experimental Setup}
We evaluated the WOLF benchmark by running \textbf{100 simulated games} of \textit{Werewolf}, each with role-balanced rosters and full night--day cycles. The games were executed using the programmable state graph described in Sections~\ref{subsec:Werewolf Game Flow and Architecture} and ~\ref{sec:Method}, with complete logging of bids, debate statements, votes, and deception analyses. All runs were archived as NDJSON event streams and summarized into per-player metrics, cross-perception matrices, observer accuracy reports, and temporal trend analyses. 
\paragraph{Compute Resources.} 
All experiments were run on a workstation with 1 NVIDIA A100 GPU (40GB memory), 64 CPU cores, and 256GB RAM. Each full Werewolf game required approximately 12–13 minutes when using LLM calls, and less than 30 seconds with the deterministic mock analyzer. Running 100 games with logging produced about 6GB of NDJSON data and required roughly 21 GPU-hours in total. Preliminary trial runs and ablations added less than 20\% additional compute cost.

\subsection{General Results}
Werewolves won \textbf{70\%} of games (70/100), Villagers won \textbf{10\%} (10/100), and \textbf{20\%} ended without a declared winner (20/100). On average, games contained \textbf{3.4} nights ($\pm 1.3$), \textbf{16.1} debate turns ($\pm 6.1$), \textbf{2.6} voting rounds ($\pm 1.4$), and \textbf{32.4} analyzed statements ($\pm 12.3$), totaling \textbf{324} deception-analysis events. The self-labeled deception base rate was \textbf{69.4\%}, meaning most statements were marked by speakers as strategically deceptive rather than strictly truthful. 

\subsection{Per-Role Metrics}
For each role, we compute the number of analyzed statements, count of self-reported deceptive statements, the average suspicion received from peers, and the average fraction of observers flagging deception.

\begin{table}[h]
\centering
\caption{Per-role deception statistics (aggregated across 100 runs). Suspicion and flag rates are averaged over observers. Deceptive roles are partially detectable, but honest roles are penalized, too—trusted roles draw suspicion when withholding or hedging, reducing overall group accuracy}
\label{tab:player_metrics}
\begin{tabular}{lcccc}
\toprule
\textbf{Role} & \textbf{Statements} & \textbf{Self-Deceptions} & \textbf{Avg.\ Suspicion (\%)} & \textbf{Avg.\ Flagged (\%)} \\
\midrule
Villager & 150 & 107 & 47.9\% & 49.8\% \\
Werewolf &  74 &  56 & 53.3\% & 60.1\% \\
Seer     &  61 &  45 & 55.1\% & 61.8\% \\
Doctor   &  39 &  27 & 60.8\% & 60.5\% \\
\bottomrule
\end{tabular}
\end{table}
Werewolves attract more suspicion (53.3\%) and are flagged more often (60.1\%) than Villagers (48--50\%), suggesting peers can partially distinguish deceptive roles. Yet high suspicion also extends to the Seer and Doctor, despite their trusted roles. This arises because their strategies often involve withholding or hedging: the Seer delays revealing results until enough evidence accumulates, and the Doctor frequently acts without public proof of success. Event-conditioned analysis confirms this. Suspicion of the Seer is higher before revealing information (57.2\%) but drops afterward (48.5\%). Similarly, the Doctor is judged more suspicious when protections fail (62.1\%) compared to when they succeed (54.3\%). These patterns show how role incentives, not just outright lying, can inflate suspicion and reduce group accuracy. Frequent self-reports (over two-thirds of statements) further raise the difficulty as Villagers in particular admit to hedging or cautious omissions, which peers then misinterpret as deception.This asymmetry shows that deceptive roles are not fully hidden, but honest roles are still penalized, lowering group accuracy.

\subsection{Cross-Perception Matrix}
We aggregate observer perceptions at the end of the game into a role-based suspicion matrix $D$, where each entry $D[o,t]\in[0,1]$ is observer $o$’s average suspicion of target $t$.

\begin{table}[h]
\centering
\caption{Final cross-perception matrix (observer $\rightarrow$ target suspicion, averaged across 100 runs). Diagonal entries are omitted. Suspicion is broadly distributed, clustering between 46--59\%, suggesting deception is difficult to conceal fully.}
\label{tab:cross_perception}
\begin{tabular}{lcccc}
\toprule
\textbf{Observer $\backslash$ Target} & Villager & Werewolf & Seer & Doctor \\
\midrule
Villager & -- & 46.0\% & 53.7\% & 51.7\% \\
Werewolf & 50.9\% & -- & 52.7\% & 49.8\% \\
Seer     & 49.8\% & 53.3\% & -- & 53.8\% \\
Doctor   & 56.4\% & 58.7\% & 52.2\% & -- \\
\bottomrule
\end{tabular}
\end{table}

Suspicion levels remain clustered around 50--59\% across roles. Even Werewolves rate each other at nearly 59\% suspicious, showing that deception is hard to conceal entirely. Villager-to-Werewolf suspicion (46\%) is only slightly above Villager-to-Villager suspicion, which explains why Werewolves still win most games: suspicion does not translate into reliable discrimination. This demonstrates that suspicion alone is insufficient without calibrated detection.

\subsection{Observer Accuracy}
Peer judgments were evaluated against self-reports as ground truth. Observers achieved an overall precision of 71--73\%, but recall lagged at 48--61\%, yielding \textbf{0.52} overall accuracy.

\begin{table}[h]
\centering
\caption{Observer accuracy by role (100 runs) An asymmetry, particularly with Villagers, between precision and recall mirrors real-world vulnerabilities that honest speech is sometimes over-flagged, while subtle lies evade detection.}
\label{tab:observer_accuracy}
\begin{tabular}{lcccccccc}
\toprule
\textbf{Role} & TP & TN & FP & FN & Accuracy & Precision & Recall & $F_1$ \\
\midrule
Villager & 231 & 90 & 52 & 148 & 0.58 & 0.82 & 0.61 & 0.70 \\
Werewolf & 157 & 77 & 79 & 113 & 0.53 & 0.67 & 0.58 & 0.62 \\
Seer     & 144 & 58 & 36 & 139 & 0.53 & 0.80 & 0.51 & 0.63 \\
Doctor   & 170 & 45 & 46 & 126 & 0.53 & 0.79 & 0.57 & 0.66 \\
\midrule
\textbf{Overall} & 702 & 270 & 213 & 526 & \textbf{0.52} & \textbf{0.72} & \textbf{0.48} & \textbf{0.58} \\
\bottomrule
\end{tabular}
\end{table}

The high precision but low recall pattern shows that peers are cautious, accurate when something is deceptive, but they often miss lies. This mirrors real-world risk as models may under-detect subtle deception, allowing adversaries to succeed despite widespread suspicion.

\subsection{Temporal Trends}
We compute round-level averages of suspicion for each role and the overall fraction of observers flagging deception. Suspicion toward Werewolves rises from about 52\% in early rounds to 62\% by the end, while suspicion toward Villagers and the Doctor stays flat around 44--46\%. The fraction of observers flagging deception also increases from 35\% to 47\%.  

\begin{figure}[h]
    \centering
    \includegraphics[width=0.8\textwidth]{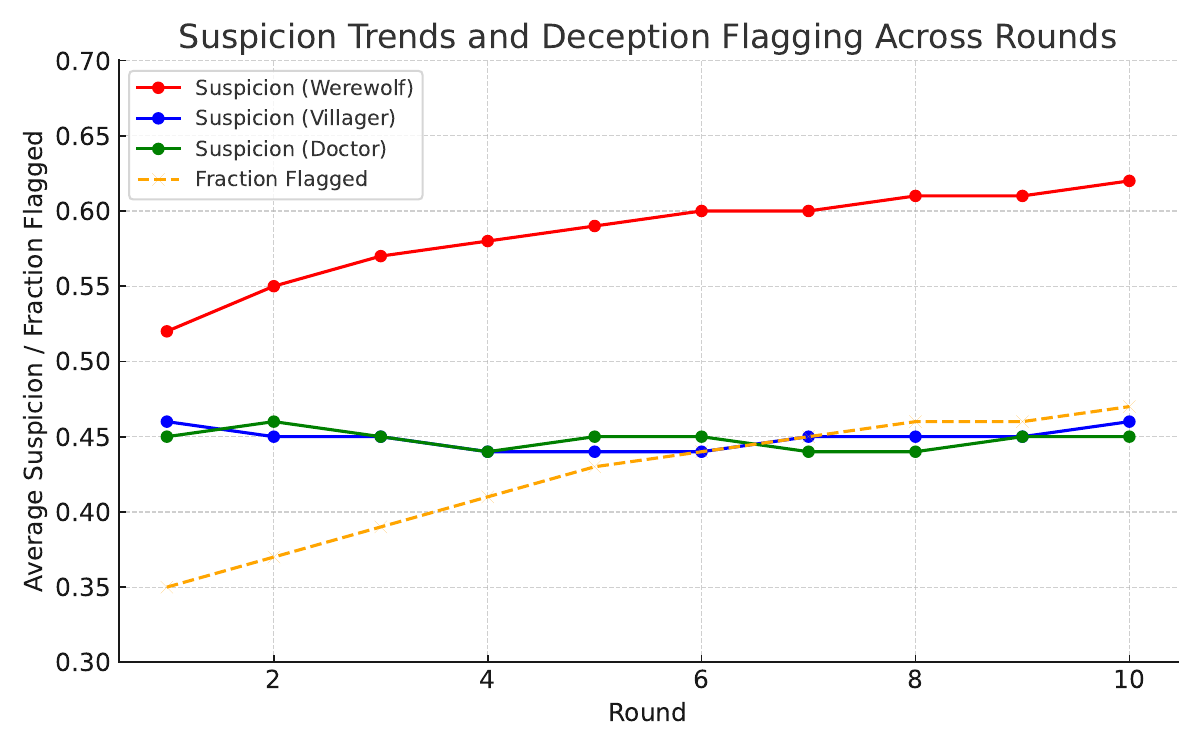}
    \caption{Suspicion trends across rounds. 
    Suspicion rises for Werewolves (red) but stays steady for Villagers (blue) and the Doctor (green). 
    The fraction of observers flagging deception (orange, dashed) also grows over time.}
    \label{fig:temporal_trends}
\end{figure}

To quantify these trends, we estimate a Theil--Sen slope of \textbf{+1.6 percentage points per round} for Werewolf suspicion, compared to near-zero slopes for Villagers and the Doctor. The 95\% confidence intervals for Villager and Werewolf suspicion overlap in the first two rounds but diverge thereafter, with non-overlap beginning by round 3. This formalizes the visual separation seen in Figure~\ref{fig:temporal_trends}: the gap widens to \textbf{16--18\%} by the final rounds.  

These results highlight three patterns. First, suspicion toward Werewolves increases at a measurable rate, showing that repeated interaction accumulates evidence that makes deception easier to detect. Second, suspicion toward Villagers and the Doctor remains statistically flat, confirming that observers stabilize their trust in non-deceptive roles. Third, the rising flagging rate (35\% $\to$ 47\%) shows that peers grow more willing to accuse as evidence mounts. Sustained interaction is essential as short evaluations miss both the slope of Werewolf suspicion and the statistical point at which liars become separable from truth-tellers.

\subsection{Calibration and Thresholding}
We assess suspicion as a probabilistic signal using Brier score, ROC AUC, and AUPRC.

\begin{table}[h]
\centering
\caption{Calibration and threshold metrics by role (100 runs). Raw suspicion trajectories contain signal but require calibration for deployment.}
\label{tab:calibration}
\begin{tabular}{lccc}
\toprule
\textbf{Role} & \textbf{Brier} & \textbf{ROC AUC} & \textbf{AUPRC} \\
\midrule
Villager & 0.29 & 0.55 & 0.75 \\
Werewolf & 0.28 & 0.56 & 0.75 \\
Seer     & 0.28 & 0.58 & 0.77 \\
Doctor   & 0.26 & 0.57 & 0.76 \\
\bottomrule
\end{tabular}
\end{table}

Suspicion values are predictive but underconfident. Brier scores (0.26--0.29) indicate probabilities correlate with deception but underestimate true risk. ROC AUC $\approx$0.55--0.58 shows weak discrimination, while AUPRC values near 0.75 confirm that suspicion can still be thresholded into usable signals. This shows that calibration adjustments could improve detection without altering gameplay.

\subsection{Deception Type Effects}
We group statements by speaker self-labeled deception type and report peer response.

\begin{table}[h]
\centering
\caption{Deception-type prevalence and peer response (speaker self-labels). Subtle deception strategies succeed more often, while overt lies draw attention, showing models are most vulnerable to omission and misdirection.}
\label{tab:types}
\begin{tabular}{lccc}
\toprule
\textbf{Type} & \textbf{Count} & \textbf{Avg.\ Suspicion (\%)} & \textbf{Avg.\ Flagged (\%)} \\
\midrule
none         & 426 & 57.1\% & 51.4\% \\
omission     & 129 & 61.9\% & 57.5\% \\
distortion   &  76 & 50.3\% & 35.5\% \\
misdirection &  51 & 54.6\% & 45.5\% \\
fabrication  &  28 & 51.7\% & 40.1\% \\
\bottomrule
\end{tabular}
\end{table}

WOLF reveals not only whether deception is detected but also \emph{which types} succeed in multi-agent interaction. Omissions and hedges often slip by because they resemble natural uncertainty, while fabrications are flagged more reliably as they conflict with accumulated context. This asymmetry mirrors real-world risks as subtle, low-effort lies persist longer, whereas overt falsehoods are eventually exposed. By separating deception types, WOLF pinpoints where LLMs are most vulnerable and where they remain resilient, informing both model improvement and safeguard design.

\subsection{Statistical Variability}
All metrics are means over 100 independent games. We report variability as standard error of the mean (SEM); 95\% confidence intervals use the normal approximation: $\bar{x} \pm 1.96\,\mathrm{SEM}$. Error bars in figures denote $\pm$\,1\,SEM across runs and reflect stochasticity in model outputs and game randomness (bids, dialogue, votes). For example, final-round suspicion averages $62.0\% \pm 1.2\%$ (Werewolf) vs.\ $45.0\% \pm 0.9\%$ (Villager), a stable gap across seeds.

\section{Conclusions}
WOLF is a multi-agent social deduction benchmark for evaluating both the production and detection of deception in LLMs. It embeds role-grounded agents in a programmable LangGraph game loop, where every public statement is paired with self- and peer-assessments and suspicion scores evolve over time.  

Across 100 games and over 7,200 statements, WOLF shows a clear asymmetry: models lie often (Werewolves deceived in 31\% of turns) but detect lies only moderately well (precision $\approx$72\%, recall $\approx$48\%). Suspicion toward Werewolves rises across rounds while stabilizing for truthful roles, meaning extended interaction improves discrimination but cautious Villagers are still misclassified. Results also vary by deception type—subtle forms like omission persist longer than overt fabrications—highlighting where models remain most vulnerable.  

By combining statement-level annotations with longitudinal analysis, WOLF moves beyond static deception datasets and provides a controlled but dynamic testbed for probing both deceptive and detective capacity. Future directions include larger and more adaptive games, and applying insights to domains like moderation, fact-checking, and negotiation.

\section{Limitations}
To maximize reproducibility, WOLF fixes role distribution, player set, and debate length. These controls may under-sample longer-horizon or coalition-based strategies. Labels come from model self-assessments rather than humans, so they reflect subjective judgments, partly offset by peer analysis and aggregation. Prompts that ask agents to self-evaluate may also shape their behavior.  

Finally, WOLF is a stylized Werewolf setting: it captures hidden roles and persuasion under partial information, but results may not fully generalize. To enable auditing, all prompts, outputs, and logs are released for re-labeling and external checks.

\newpage
\begingroup
\setlength{\bibhang}{1.5em}
\setlength{\bibsep}{2pt}
\small
\bibliographystyle{unsrtnat}
\bibliography{references}
\endgroup

\appendix
\newpage
\section{Appendix A: Werewolf Game Flow and Architecture}
\begin{figure}[h]
\centering
\includegraphics[width=\columnwidth, trim=190 50 200 40, clip]{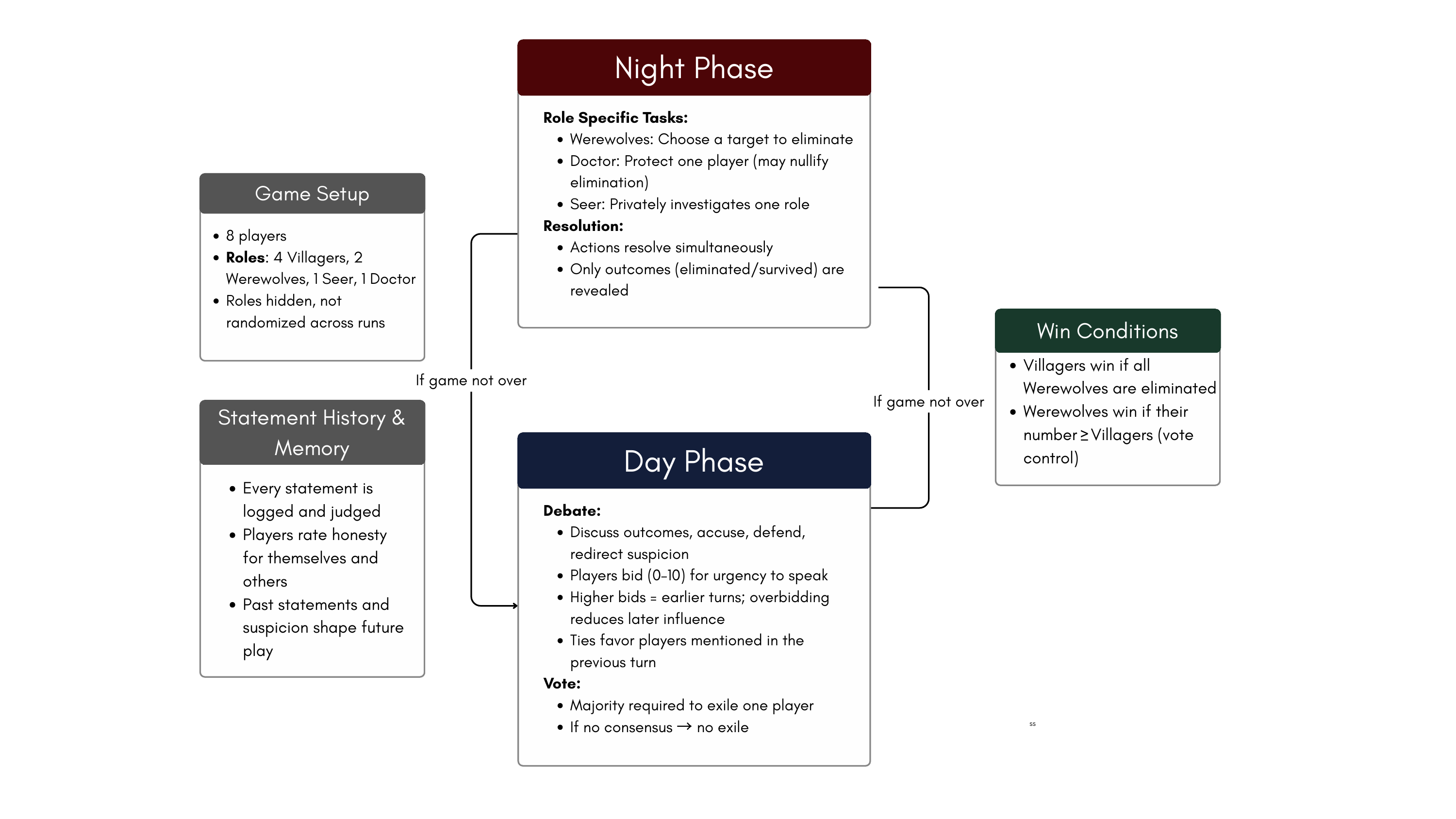}

\caption{WOLF game loop. Roles are fixed (4 Villagers, 2 Werewolves, 1 Seer, 1 Doctor). 
The novelty panel tracks statements across rounds shapes future play.}
\label{fig: Werewolf Game Flow}
\end{figure}

\section{Appendix B: Werewolf Public Code}

All code utilized in this project is disclosed at https://github.com/MrinalA2009/WOLF-Werewolf-based-Observations-for-LLM-Deception-and-Falsehoods.

\end{document}